\documentclass[aps,prl,showpacs,twocolumn,superscriptaddress]{revtex4-1}
\usepackage{amsmath}
\usepackage{amsfonts}
\usepackage{graphicx}
\usepackage{dcolumn}
\usepackage{amsbsy}
\usepackage{bm}
\usepackage{amsthm}
\usepackage[caption=false]{subfig}
\usepackage{tabularx}
\usepackage{array}
\usepackage{natbib}
\usepackage{hyperref}

\begin{document}
\title{Hawking Effect as Quantum Inertial Effect}

\author{Yu-Lei Feng}
\email{11336008@zju.edu.cn}
 \affiliation{Zhejiang Institute of Modern Physics, Zhejiang University, Hangzhou 310027, China}
\author{Yi-Xin Chen}
 \email{yixinchenzimp@zju.edu.cn}
 \affiliation{Zhejiang Institute of Modern Physics, Zhejiang University, Hangzhou 310027, China}

\begin{abstract}
  We show that ``particle production" by gravitational field, especially the Hawking effect, may be treated as some quantum inertial effect, with the energy of Hawking radiation as some vacuum energy shift. This quantum inertial effect is mainly resulted from some intrinsical energy fluctuation $\hbar\kappa/c$ for a black hole. In particular, there is an extreme case in which $\hbar\kappa/c$ is the Planck energy, giving a ``Planck black hole" whose event horizon's diameter is one Planck length. Moreover, we also provide a possibility to obtain some positive cosmological constant for an expanding universe, which is induced from the vacuum energy shift caused by quantum inertial effect.
\end{abstract}
\maketitle

\emph{Introduction}.---Hawking effect~\cite{a,b}, which is believed to cause black hole evaporation, is a mysterious feature of quantum fields in a curved spacetime. It belongs to a more general effect of ``particle production" by gravitational field. However, the spectrum of the ``produced particles" seems to be thermal, leading to the information loss paradox~\cite{b} for a black hole.
Another firewall paradox~\cite{c,d,e,f,f1,g,g1} was proposed by Almheiri et al. It roughly says the near horizon region will become a firewall if black hole evaporation is unitary.

These paradoxes indicate that \emph{Hawking effect may not be able to cause black hole evaporation}. This argument can roughly be confirmed by an effective unitary model of black hole evaporation proposed in our papers~\cite{h,i}. In that model, the entangled in-falling vacuum is used as some medium to induce nonlocal correlations between the black hole interior and exterior.

In this paper, we give another interpretation of Hawking effect as some quantum inertial effect, in the framework of effective field theory. To understand it, let's first consider the classical inertial effect.

\begin{figure}[tbp]
\setlength{\unitlength}{1mm} \centering
\includegraphics[width=2.4in]{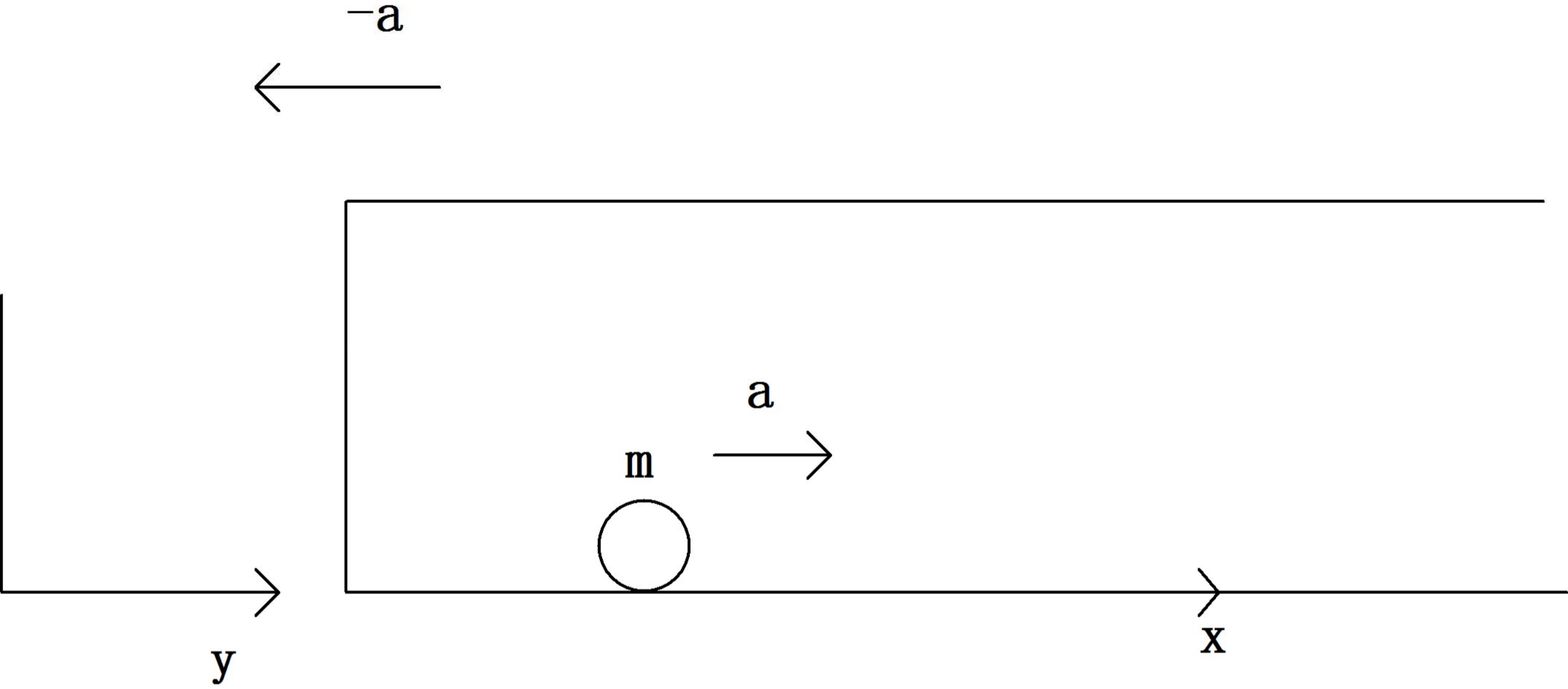}
\caption{\label{fig:1} Inertial force effect. }
\end{figure}

\emph{Classical inertial effect}.---Consider a semi-infinite carriage containing a smooth ball with a mass $m$ sitting on the smooth floor, see figure~\ref{fig:1}.
Supposing both the carriage and ball are still initially, then let the carriage start with a constant acceleration ``$-a$". As is well known, the ball will move with an acceleration ``$a$" respect to an observer fixed on the carriage, with coordinate $(x^0,x^1)$. In other words, there seems to be some \emph{inertial force} driving the ball to move with its energy changed as
\begin{equation}
\label{eq:x}
\begin{split}
E_0=0\longrightarrow E=\frac{1}{2}m(ax^0)^2\propto a^2
\,.
\end{split}
\end{equation}
However, to another observer outside the carriage with coordinate $(y^0,y^1)$, the ball will always be still due to its inertia. This is because that the frame $(x^0,x^1)$ is non-inertial, which is related to the inertial frame $(y^0,y^1)$ through the following transformations
\begin{equation}
\label{eq:y}
\begin{split}
y^0=x^0,~~y^1=x^1-\frac{1}{2}a(x^0)^2
\,.
\end{split}
\end{equation}
Further, if the ball has a constant velocity $v_0$ initially, then after the carriage starts, its energy respect to the frame $(x^0,x^1)$ will be
\begin{equation}
\label{eq:z}
\begin{split}
E=\frac{1}{2}mv^2_0+mv_0ax^0+\frac{1}{2}m(ax^0)^2
\,.
\end{split}
\end{equation}
In addition to the ``$\propto a^2$" term, there is also a ``$\propto a$" contribution. By comparing with Eq.~\eqref{eq:x}, we can see that the ``$\propto a^2$" term may be treated as some ``vacuum energy shift", with the original ``vacuum energy" $E_0=0$.

The inertial force exerted on the ball, can also be treated to be from some gravitational field locally, according to Einstein's equivalence principle~\cite{j}. In other words, there is an effective transformation or evolution from a flat metric to a curved one
\begin{equation}
\label{eq:z1}
\begin{split}
\eta_{\mu\nu}(x)\longrightarrow g_{\mu\nu}(x)
\,,
\end{split}
\end{equation}
whose effect is equivalent to that of the coordinate transformation in Eq.~\eqref{eq:y} between inertial and non-inertial frames.
These can be described by \emph{diffeomorphism}~\cite{j}, with the metric or more general tensor transformation as an active viewpoint and coordinate transformation as a passive viewpoint.

\emph{Hawking effect as quantum inertial effect}.---Let's consider effective field theory in a curved spacetime, such as a black hole $g^B_{\mu\nu}$ under formation. In addition to Eq.~\eqref{eq:z1}, there is also a transformation for the energy-momentum tensor operator for some quantum (scalar) field $\hat{\phi}$
\begin{equation}
\label{eq:z2}
\begin{split}
\hat{T}_{\mu\nu}[\eta,\hat{\phi}](x)\stackrel{\hat{U}}{\longrightarrow}\hat{T}_{\mu\nu}[g^B,\hat{\phi}](x)
\,.
\end{split}
\end{equation}
This evolution is unitary but incompletely due to the classical background.
In a passive viewpoint, there are corresponding transformations between a static frame and an in-falling frame.

It's the expectation value $\langle \hat{T}_{\mu\nu}\rangle$ that appears in semiclassical Einstein's equation~\cite{k}.
Note that quantum state is \emph{fixed} in the Heisenberg picture, and it is background or coordinate frame dependent. Actually, vacuum can only be found for some specific background with \emph{isometry} symmetries, for example Lorentz symmetry for a flat metric. This is because (time-like) Killing vector of an isometry symmetry is crucial to define a vacuum~\cite{l1}. Then we have
\begin{equation}
\label{eq:z5}
\begin{split}
``object":\arrowvert0_\eta\rangle,~~``observable":\hat{T}_{\mu\nu}[g^B,\hat{\phi}](x)
\,,
\end{split}
\end{equation}
where the (initial) vacuum is like the ball in the classical inertial effect, and the observable is like the observer fixed on the carriage. In this sense, \emph{the (initial) quantum state seems to possess some ``inertia" associated with the chosen background or coordinate frame}. Then, \emph{``particle production" by gravitational field, especially the Hawking effect, may be treated as some quantum inertial effect}, due to some ``inertial force" or background gravity ``acted" on the (initial) quantum state.

Note that inertial effect mainly stands for the influences of inertial force or background gravity on the observations.
It thus seems that the above quantum inertial effect \emph{cannot} be treated simply as the back-reaction of matter via $\langle \hat{T}_{\mu\nu}\rangle$, otherwise general covariance of Einstein's equation would be broken. For example, Hawking effect can occur in a static frame via $\langle0_\eta\arrowvert\hat{T}_{\mu\nu}[g^B]\arrowvert0_\eta\rangle$, but it does not happen in an in-falling frame, since the expectation value is $\langle0_\eta\arrowvert\hat{T}_{\alpha\beta}[\eta]\arrowvert0_\eta\rangle$. If Hawking effect can cause a black hole evaporation as the back-reaction of matter, then what about the description for the in-falling frame? This problem can also be seen as follows.

Actually, $\langle0_\eta\arrowvert\hat{T}_{\mu\nu}[g^B]\arrowvert0_\eta\rangle$ has the same effect only under isometry transformations of the black hole. Now, the semiclassical Einstein's equation is given by
\begin{equation}
\label{eq:z6}
\begin{split}
G_{\mu\nu}[g^B+h]=8\pi G\langle0_\eta\arrowvert\hat{T}_{\mu\nu}[g^B]\arrowvert0_\eta\rangle,
\,.
\end{split}
\end{equation}
where metric in Einstein tensor $G_{\mu\nu}$ is perturbed due to the back-reaction. This means that \emph{the isometry symmetry for $g^B_{\mu\nu}$ would be broken, if $g^B_{\mu\nu}+h_{\mu\nu}$ was another stationary black hole with some different isometry symmetry}. In this sense, $h_{\mu\nu}$ can only be some small metric perturbation, provided semiclassical Einstein's equation is applicable. This implies that black hole evaporation, from stationary background to another, cannot be caused by the Hawking effect, confirming that Hawking effect is some quantum inertial effect.

Since ``particle production" by gravitational field can be treated as quantum inertial effect, the energy of those ``produced particles" is thus the work done by the ``inertial force" or background gravity. Notice further that the energy for the Hawking~\cite{a} or Unruh effect~\cite{k1} is also of the form $\propto \kappa^2$ or $\propto a^2$, just like Eq.~\eqref{eq:x}. This indicates that the Hawking or Unruh effect can lead to some vacuum energy shift. Let's see it in some detail.

\emph{Vacuum energy shift}.---For simplicity, we assume that Hamiltonian can be constructed for a free massless scalar field in a Schwarzschild black hole. The initial and final field $\hat{\phi}^I=\hat{\phi}+\hat{\tilde{\phi}}$ can be expanded formally in terms of s-wave components~\cite{k2}
\begin{equation}
\label{eq:a}
\begin{split}
\int_{0}^{\infty}
d\omega(\hat{a}_{\omega}U_{\omega}+h.c.) = \int_{0}^{\infty}
d\omega(\hat{b}_{\omega}u_{\omega}+\hat{\tilde{b}}_{\omega}\tilde{u}_{\omega}+h.c.)
\,,
\end{split}
\end{equation}
with $U_{\omega}$ and $u_{\omega},\tilde{u}_{\omega}$ the modes at the infinite past $I^-$, the infinite future $I^+$ and the future horizon $H^+$ respectively.

At the infinite future $I^+$, the expectation value of the Hamiltonian for initial Minkowski vacuum is
\begin{equation}
\label{eq:b}
\begin{split}
\langle0_I\arrowvert \hat{H}\arrowvert0_I\rangle=\int_{0}^{\infty}\frac{d\omega}{2\pi}(\hbar\omega)\langle0_I\arrowvert[\hat{b}^{\dag}_{\omega}\hat{b}_{\omega}+\frac{1}{2}T]\arrowvert0_I\rangle,
\end{split}
\end{equation}
where we retain the infinite zero point energy, and introduce a factor $T\equiv\lim_{\omega_1\rightarrow\omega_2}\delta(\omega_1-\omega_2)$, formally treated as a long period of time. Certainly, this factor may be avoided by introducing discrete quantum numbers~\cite{k2}.
After some calculations, we have
\begin{equation}
\label{eq:c}
\begin{split}
E^{\phi}_{vac}&=T\int_{0}^{\infty}\frac{d\omega}{2\pi}(\hbar\omega)[\frac{1}{2}+\frac{1}{e^{2\pi c\omega/\kappa}-1}]\\
&=E^{\phi}_{\infty}+D\hbar(\frac{\kappa}{c})^2,
\end{split}
\end{equation}
where $D$ is some constant, and $(e^{2\pi c\omega/\kappa}-1)^{-1}$ is the Hawking factor, with the surface gravity $\kappa=c^4/4GM$. Easily to see, in addition to the ordinary infinite zero point energy $E^{\phi}_{\infty}$, there is also a finite term ``$\propto\kappa^{2}$" up to the factor $T$, just like the shift in Eq.~\eqref{eq:x}.

Now, let's consider another expectation value for the (normalized) first excited state $\langle0_I\arrowvert \hat{a}_{\omega_0} \hat{H}\hat{a}^{\dag}_{\omega_0}\arrowvert0_I\rangle/T$.
After some tedious calculations, we have
\begin{equation}
\label{eq:e}
\begin{split}
E^{\phi}_{vac}+\frac{1}{T}\int_{0}^{\infty}\frac{d\omega}{2\pi}(\hbar\omega)[|\alpha_{\omega\omega_0}|^2+|\beta_{\omega\omega_0}|^2],
\end{split}
\end{equation}
where $\alpha$ and $\beta$ are corresponding Bogolubov coefficients. By using of the relation $|\alpha_{\omega\omega_0}|=e^{\pi c\omega/\kappa}|\beta_{\omega\omega_0}|$ and the expression for $|\beta_{\omega\omega_0}|^2$~\cite{k3}
\begin{equation}
\label{eq:f}
\begin{split}
|\beta_{\omega\omega_0}|^2=\frac{c/\kappa}{2\pi\omega_0}\frac{1}{e^{2\pi c\omega/\kappa}-1},
\end{split}
\end{equation}
the second term in Eq.~\eqref{eq:e} will become
\begin{equation}
\label{eq:f1}
\begin{split}
\frac{c/\kappa}{\pi\omega_0T}\int_{0}^{\infty}\frac{d\omega}{2\pi}(\hbar\omega)[\frac{1}{2}+\frac{1}{e^{2\pi c\omega/\kappa}-1}].
\end{split}
\end{equation}
Then the total energy for the excited state is
\begin{equation}
\label{eq:g}
\begin{split}
E^{\phi}_{\infty}+D\hbar(\frac{\kappa}{c})^2+E_{\omega_0}^{\phi}+D'_{\omega_0}\hbar\frac{\kappa}{c},
\end{split}
\end{equation}
where $E_{\omega_0}^{\phi}$ is denoted as the first term in Eq.~\eqref{eq:f1}, and $D'_{\omega_0}$ is another constant depending on $\omega_0$. Although the form of $E_{\omega_0}^{\phi}$ is strange, the whole expression in Eq.~\eqref{eq:g} is roughly parallelled with that of Eq.~\eqref{eq:z}. Similar analysis can be made for more excited states.

Hence, for a (massless) scalar field in a (Schwarzschild) black hole, there is some finite vacuum energy shift with a form ``$\propto\kappa^2$", above the ordinary infinite zero point energy, as shown in Eq.~\eqref{eq:c}. Certainly, this shift can be treated to be caused by some ``inertial force" or background gravity on a semiclassical level. Then, what's the fundamental source of quantum inertial effect? It needs some deeper interpretation of $\hbar\kappa/c$.

\emph{Interpretation of $\hbar\kappa/c$: a first glance}.---In Hawking's viewpoint, $\hbar\kappa/c$ is related to the thermal energy $k_BT$, and the surface gravity $\kappa$ is treated as the ``temperature" of a black hole~\cite{a}. In our viewpoint, Hawking effect is a quantum inertial effect, then how to interpret $\hbar\kappa/c$?

\begin{figure}[tbp]
\setlength{\unitlength}{1mm} \centering
\includegraphics[width=3.2in]{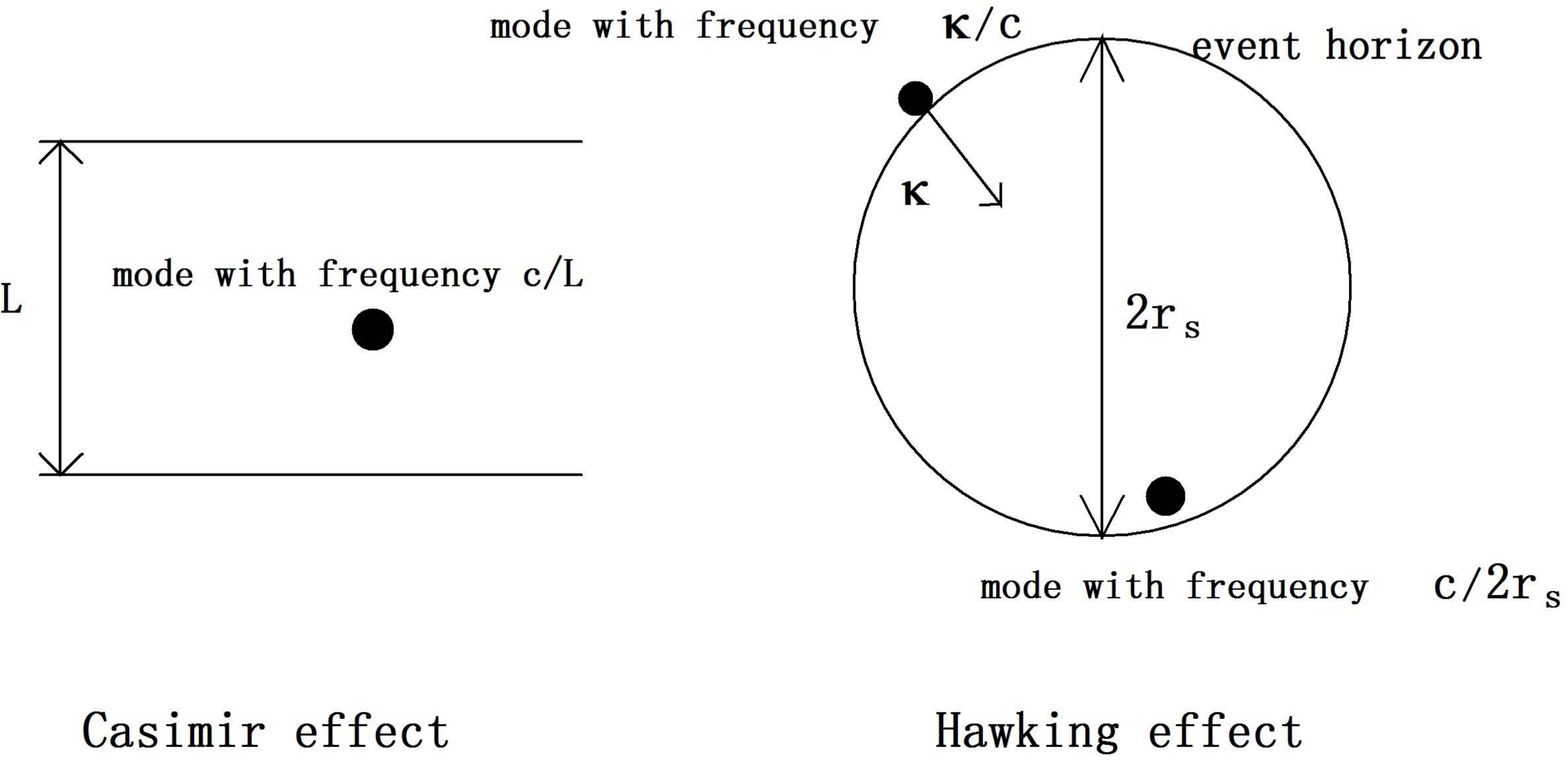}
\caption{\label{fig:2} Casimir effect and Hawking effect.}
\end{figure}

It's useful to compare with Casimir effect~\cite{l}, which also leads to some vacuum energy shift above infinite zero point energy. In that case, there are modes with energy quantum $\hbar c/L$ for standing waves perpendicular to two infinite parallel conducting plates, with $L$ the distant between the two plates, see figure~\ref{fig:2}.

For Hawking effect, it's convenient to consider a (non-relativistic) classical model first. Suppose that a particle is in a circular orbit around a large object with a mass $M$. Then the particle has a Newton acceleration $a=GM/r^2$, and its circular frequency is given by
\begin{equation}
\label{eq:7}
\begin{split}
\omega=a/v,~~v=r\omega.
\end{split}
\end{equation}
Substituting the velocity, we have
\begin{equation}
\label{eq:7a}
\begin{split}
\omega=\sqrt{\frac{a}{r}}=\sqrt{\frac{GM}{r^3}}\stackrel{r=r_S}{\longrightarrow}\sqrt{2}\frac{c^3}{4GM}=\sqrt{2}\frac{\kappa}{c},
\end{split}
\end{equation}
where the Schwarzschild radius $r_S=2GM/c^2$ has been substituted \emph{formally}. Certainly, this derivation is not rigorous, and a relativistic analysis for a black hole is more complicated~\cite{j}. From Eq.~\eqref{eq:7a}, we can see that $\kappa/c$ is \emph{like} a unit of circular frequency for particles in circular orbits around event horizon. Although the derivation is only \emph{pro forma}, it indicates that $\hbar\kappa/c$ seems to be some ``energy quantum" for modes moving around event horizon, as shown in figure~\ref{fig:2}. Certainly, this ``energy quantum" is mainly caused by the surface gravity or ``inertial force" on a semiclassical level.

This interpretation is suited to (massless) fields within both the black hole exterior and interior.
For fields within a Schwarzschild black hole interior, there is also another \emph{rough} interpretation similar to that of Casimir effect. Note that the event horizon is like the two conducting plates for Casimir effect, with a distant given by $2r_S$. Then the ``energy quantum" will be
\begin{equation}
\label{eq:7b}
\begin{split}
\hbar c/2r_S=\hbar c^3/4GM=\hbar\kappa/c,
\end{split}
\end{equation}
which is also shown in figure~\ref{fig:2}.

Differing from Casimir effect in which $\hbar c/L$ is a \emph{real} energy quantum of the confined field, $\hbar\kappa/c$ for Hawking effect seems to be a ``cutoff" via the Hawking factor
\begin{equation}
\label{eq:7c}
\begin{split}
(e^{2\pi\frac{\hbar\omega}{\hbar\kappa/c}}-1)^{-1}\simeq e^{-2\pi\frac{\hbar\omega}{\hbar\kappa/c}},~ (\hbar\omega\gg\hbar\kappa/c).
\end{split}
\end{equation}
That is, contributions from energy levels (largely) higher than $\hbar\kappa/c$ will be suppressed by a damping factor. This difference is mainly because that Hawking effect is a quantum inertial effect, with $\hbar\kappa/c$ as the ``energy quantum" caused by the surface gravity or ``inertial force". Then, energy levels higher than $\hbar\kappa/c$ can not be excited easily, and the contributions of the vacuum energy shift are mainly from energy levels lower than $\hbar\kappa/c$.

\emph{$\hbar\kappa/c$ as energy fluctuation of a black hole}.---The above interpretation of $\hbar\kappa/c$ is not fundamental, since it's derived only through analogy on a semiclassical level. Note that surface gravity comes from black hole, it indicates that $\hbar\kappa/c$ may be an \emph{intrinsical} feature of a black hole.

Actually, $c/\kappa$ provides a \emph{time scale} for physics happening in a (Schwarzschild) black hole. This can be seen more clearly by collecting a class of quantities
\begin{equation}
\label{eq:7c1}
\begin{split}
(2r_S,~c/\kappa,~M),
\end{split}
\end{equation}
which provides the scales of length, time and mass (or energy via $Mc^2$).
In Planck unites~\cite{j}, we can set
\begin{equation}
\label{eq:7d}
\begin{split}
\gamma 2r_S=nl_P,~(n=1,2,\cdots),~~l_P=(\frac{\hbar G}{c^3})^{1/2},
\end{split}
\end{equation}
with $l_P$ the Planck length, and $\gamma$ some positive dimensionless constant that may be absorbed into the definition of $l_P$. Then, the black hole's mass is given by
\begin{equation}
\label{eq:7e}
\begin{split}
M=\frac{n}{4\gamma}m_P,~~m_P=(\frac{\hbar c}{G})^{1/2},
\end{split}
\end{equation}
with $m_P$ the Planck mass. And $c/\kappa$ is given by
\begin{equation}
\label{eq:7e1}
\begin{split}
c/\kappa=\frac{n}{\gamma}t_P,~~t_P=(\frac{\hbar G}{c^5})^{1/2},
\end{split}
\end{equation}
with $t_P$ the Planck time. Analogous to the fundamental scales $(l_P,t_P,m_P)$, $(2r_S,\kappa/c,M)$ provides some intermediate scales for a black hole. Certainly, when the black hole is evolutive, these scales will also vary.

According to uncertainty principle, $\hbar\kappa/c$ is a measure of \emph{energy uncertainty or fluctuation} for physics happening in a black hole, including the black hole itself. When $\hbar\kappa/c\ll Mc^2$, the black hole is stable enough, and the scales in Eq.~\eqref{eq:7c1} are appropriate, especially the time scale $c/\kappa$ can be used to provide redshift factor $e^{-\frac{u}{c/\kappa}}$ for outgoing radiation~\cite{k2}. However, when $\hbar\kappa/c\lesssim Mc^2$, the black hole is unstable and the scales in Eq.~\eqref{eq:7c1} will be useless, leaving only the fundamental scales $(l_P,t_P,m_P)$.
Actually, in Planck unites, $\hbar\kappa/c$ is given by
\begin{equation}
\label{eq:7f}
\begin{split}
\frac{\hbar\kappa}{c}=\frac{\gamma}{n}E_P,~~E_P=c^2(\frac{\hbar c}{G})^{1/2},
\end{split}
\end{equation}
with $E_P$ the Planck energy. Easily to see, the energy fluctuation roughly has the same order with black hole's energy for \emph{micro} black holes. Besides, there is an upper bound for $\hbar\kappa/c$, the Planck energy.

There is an extreme case when $\gamma=1,n=1$
\begin{equation}
\label{eq:7h}
\begin{split}
2r_S=l_P,~c/\kappa=t_P,~M=m_P/4,~\hbar\kappa/c=E_P,
\end{split}
\end{equation}
which can be called a ``Planck black hole" with the lowest energy, since the diameter of its event horizon is given by one Planck length. Note that flat space cannot exist in a spacetime containing matter. Thus, this ``Planck black hole" may be treated as some ``\emph{vacuum}" of quantum spacetime. Certainly, this black hole cannot be expressed in terms of classical metric, since the spacetime is discrete in terms of fundamental scales, thus the singularity at $r=0$ is avoided.

The energy fluctuation $\hbar\kappa/c$ also serves as a measure for the reliability of effective field theory. For a large black hole with $\hbar\kappa/c\ll Mc^2$, it's stable enough so that effective field theory is reliable, with the back-reaction only as small metric perturbations. Besides, quantum inertial or Hawking effect may occur, giving vacuum energy shift with contributions mainly from energy levels lower than $\hbar\kappa/c$. In this sense, \emph{it's the black hole's intrinsical energy fluctuation that provides the fundamental source of quantum inertial effect}, giving ``energy quantum" $\hbar\kappa/c$ for modes moving around the
event horizon on a semiclassical level. For a moderate black hole with $\hbar\kappa/c\lesssim Mc^2$, it's \emph{not} stable enough so that effective field theory can only give some qualitative descriptions, and a quantum gravity theory is needed.

\emph{Expanding universe case}.---Quantum inertial effect can also occur for an evolutive universe. Consider an expanding universe described by the FRW metric~\cite{n}
\begin{equation}
\label{eq:8}
\begin{split}
ds^2=-dt^2+a(t)^2[\frac{dr^2}{1-Kr^2}+r^2d\Omega^2].
\end{split}
\end{equation}
For simplicity, we consider a simple model with $K=0$, with the scale factor satisfying~\cite{m,o}
\begin{equation}
\label{eq:9}
\begin{split}
a_1\stackrel{t\rightarrow-\infty}{\longleftarrow}a(t)\stackrel{t\rightarrow+\infty}{\longrightarrow}a_2, ~(a_1<a_2).
\end{split}
\end{equation}
Then for a scalar field, its vacuum energy at $t\rightarrow+\infty$, averaged over the initial vacuum $\arrowvert0_{-}\rangle$ is given by~\cite{m}
\begin{equation}
\label{eq:10}
\begin{split}
\sum_{\vec{k}}\hbar\omega_k(|\beta_k|^2+\frac{1}{2}),
\end{split}
\end{equation}
with $\beta_k$ the corresponding Bogolubov coefficient. This may give a finite vacuum energy shift above the ordinary infinite zero point energy, similarly for the case $K\neq0$.

As shown above, the fundamental source of quantum inertial effect is some intrinsical energy fluctuation of the background. For an expanding universe, this fluctuation is mainly caused by the change of the scale factor $a(t)$, thus we can give an estimate $\hbar \dot{a}/a$, with $\dot{a}$ the time derivative of $a(t)$. In particular, for a de Sitter space with $a(t)\sim e^{Hct}$, the energy fluctuation is $\hbar Hc$. This can be confirmed by the Gibbons-Hawking effect for a de Sitter space, with a surface gravity given by $\kappa_{de}\simeq c^2H$~\cite{q1}.

Then, how does this quantum inertial effect influence the universe's evolution? One may directly replace the classical energy-momentum tensor by $\langle \hat{T}_{\mu\nu}\rangle$~\cite{o1,o2}. However, this will break the general covariance of Einstein's equation, as shown around Eq.~\eqref{eq:z6}.
In fact, quantum inertial effect can lead to some vacuum energy shift. This energy shift can be included in an approximate but covariant way~\cite{p,p1}, by modifying the potential of some given field for an evolution \emph{in progress}
\begin{equation}
\label{eq:11}
\begin{split}
-V(\phi)g_{\mu\nu}\longrightarrow-(V(\phi)+\delta\mathcal{E}^{\phi}_{vac})g_{\mu\nu},
\end{split}
\end{equation}
with $\delta\mathcal{E}^{\phi}_{vac}$ the density of vacuum energy shift resulted from \emph{previous} evolutions. This modification may induce some positive cosmological constant.

An estimate of density of the vacuum energy shift in Eq.~\eqref{eq:10} is $\sim \hbar ca^{-4}$~\cite{q}, whose effect is negligible in the early era due to the smallness of $\hbar$. However, during universe's expansion, the amount of matter and radiation is roughly unchanged, with their densities decreasing. While the amount of vacuum energy shift is \emph{accumulated}, though its density may not increase significantly. Thus, there is a stage when the density of vacuum energy shift dominates, then the universe will become de-Sitter space-like. A detail investigation is still needed.

\emph{Conclusions}.---The effect of ``particle production" by gravitational field, especially the Hawking effect, can be treated as quantum inertial effect. This inertial effect is mainly resulted from the intrinsical energy fluctuation of some specific curved background, for example $\hbar\kappa/c$ for a black hole. Besides, this inertial effect can lead to some vacuum energy shift, which may induce some positive cosmological constant for an expanding universe.

\acknowledgments
This work is supported by the NNSF of China, Grant No. 11375150.

\end{document}